\documentclass[%
 reprint,
 superscriptaddress,
 amsmath,amssymb,
 aps,
]{revtex4-2}

\usepackage{graphicx}% Include figure files
\usepackage{dcolumn}% Align table columns on decimal point
\usepackage{bm}% bold math
\usepackage{appendix}
\usepackage{hyperref}

\begin{document}

\preprint{APS/123-QED}

%PACS: in nuclear scattering, 25.30.Pt     interactions, 13.15.+g
\title{Measurement of ${}^{nat}$Pb($\nu_e$,X$n$) production with a stopped-pion neutrino source}% Force line breaks with \\

%\collaboration{COHERENT Collaboration}%\noaffiliation

\date{\today}

\newcommand{\Dukedesc}{\affiliation{Department of Physics, Duke University, Durham, NC, 27708, USA}}
\newcommand{\TUNLdesc}{\affiliation{Triangle Universities Nuclear Laboratory, Durham, NC, 27708, USA}}
\newcommand{\UTKdesc}{\affiliation{Department of Physics and Astronomy, University of Tennessee, Knoxville, TN, 37996, USA}}
\newcommand{\Mephidesc}{\affiliation{National Research Nuclear University MEPhI (Moscow Engineering Physics Institute), Moscow, 115409, Russian Federation}}
\newcommand{\ITEPnewadesc}{\affiliation{National Research Center  ``Kurchatov Institute'' , Moscow, 123182, Russian Federation }}
\newcommand{\USDdesc}{\affiliation{Department of Physics, University of South Dakota, Vermillion, SD, 57069, USA}}
\newcommand{\NCCUdesc}{\affiliation{Department of Mathematics and Physics, North Carolina Central University, Durham, NC, 27707, USA}}
\newcommand{\NCSUdesc}{\affiliation{Department of Physics, North Carolina State University, Raleigh, NC, 27695, USA}}
\newcommand{\Sandiadesc}{\affiliation{Sandia National Laboratories, Livermore, CA, 94550, USA}}
\newcommand{\ORNLdesc}{\affiliation{Oak Ridge National Laboratory, Oak Ridge, TN, 37831, USA}}
\newcommand{\UWdesc}{\affiliation{Center for Experimental Nuclear Physics and Astrophysics \& Department of Physics, University of Washington, Seattle, WA, 98195, USA}}
\newcommand{\LANLdesc}{\affiliation{Los Alamos National Laboratory, Los Alamos, NM, 87545, USA}}
\newcommand{\Laurentiandesc}{\affiliation{Department of Physics, Laurentian University, Sudbury, Ontario, P3E 2C6, Canada}}
\newcommand{\IUdesc}{\affiliation{Department of Physics, Indiana University, Bloomington, IN, 47405, USA}}
\newcommand{\VTdesc}{\affiliation{Center for Neutrino Physics, Virginia Tech, Blacksburg, VA, 24061, USA}}
\newcommand{\NCSUnucengdesc}{\affiliation{Department of Nuclear Engineering, North Carolina State University, Raleigh, NC, 27695, USA}}
\newcommand{\CMUdesc}{\affiliation{Department of Physics, Carnegie Mellon University, Pittsburgh, PA, 15213, USA}}
\newcommand{\UFdesc}{\affiliation{Department of Physics, University of Florida, Gainesville, FL, 32611, USA}}
\newcommand{\Tuftsdesc}{\affiliation{Department of Physics and Astronomy, Tufts University, Medford, MA, 02155, USA}}
\newcommand{\Concorddesc}{\affiliation{Department of Physical and Environmental Sciences, Concord University, Athens, WV, 24712, USA}}
\newcommand{\SNUdesc}{\affiliation{Department of Physics and Astronomy, Seoul National University, Seoul, 08826, Korea}}
\author{P.~An}\Dukedesc\TUNLdesc
\author{C.~Awe}\Dukedesc\TUNLdesc
\author{P.S.~Barbeau}\Dukedesc\TUNLdesc
\author{B.~Becker}\UTKdesc
\author{S.W.~Belling}\altaffiliation{Also at: Department of Electrical and Computer Engineering, University of Wisconsin-Madison, Madison, WI, 53706, USA}\TUNLdesc
\author{V.~Belov }\Mephidesc\ITEPnewadesc
\author{I.~Bernardi}\UTKdesc
\author{C.~Bock}\USDdesc
\author{A.~Bolozdynya}\Mephidesc
\author{R.~Bouabid}\Dukedesc\TUNLdesc
\author{A.~Brown }\NCCUdesc\TUNLdesc
\author{J.~Browning}\NCSUdesc
\author{B.~Cabrera-Palmer}\Sandiadesc
\author{M.~Cervantes}\Dukedesc
\author{E.~Conley}\Dukedesc
\author{J.~Daughhetee}\ORNLdesc
\author{J.~Detwiler}\UWdesc
\author{K.~Ding}\USDdesc
\author{M.R.~Durand}\UWdesc
\author{Y.~Efremenko}\UTKdesc\ORNLdesc
\author{S.R.~Elliott}\LANLdesc
\author{L.~Fabris}\ORNLdesc
\author{M.~Febbraro}\ORNLdesc
\author{A.~Gallo Rosso}\Laurentiandesc
\author{A.~Galindo-Uribarri}\ORNLdesc\UTKdesc
\author{M.P.~Green}\TUNLdesc\ORNLdesc\NCSUdesc
\author{J.~Hakenm{\"u}ller}\Dukedesc
\author{M.R.~Heath}\ORNLdesc
\author{S.~Hedges}\email{hedges3@llnl.gov}\altaffiliation{Also at: Lawrence Livermore National Laboratory, Livermore, CA, 94550, USA}\Dukedesc\TUNLdesc
\author{M.~Hughes}\IUdesc
\author{B.A.~Johnson}\IUdesc
\author{T.~Johnson}\Dukedesc\TUNLdesc
\author{A.~Khromov}\Mephidesc
\author{A.~Konovalov}\altaffiliation{Also at: Lebedev Physical Institute of the Russian Academy of Sciences, Moscow, 119991, Russian Federation}\Mephidesc
\author{E.~Kozlova}\Mephidesc
\author{A.~Kumpan}\Mephidesc
\author{O.~Kyzylova}\VTdesc
\author{L.~Li}\Dukedesc\TUNLdesc
\author{J.M.~Link}\VTdesc
\author{J.~Liu}\USDdesc
\author{A.~Major}\Dukedesc
\author{K.~Mann}\NCSUdesc
\author{D.M.~Markoff}\NCCUdesc\TUNLdesc
\author{J.~Mastroberti}\IUdesc
\author{J.~Mattingly}\NCSUnucengdesc
\author{K.~Miller}\altaffiliation{Also at: University of Chicago, Chicago, IL, 60637, USA}\Dukedesc\TUNLdesc
\author{P.E.~Mueller}\ORNLdesc
\author{J.~Newby}\ORNLdesc
\author{D.S.~Parno}\CMUdesc
\author{S.I.~Penttila}\ORNLdesc
\author{D.~Pershey}\Dukedesc
\author{C.G.~Prior}\Dukedesc\TUNLdesc
\author{R.~Rapp}\altaffiliation{Also at: Washington \& Jefferson College, Washington, PA, 15301, USA}\CMUdesc
\author{H.~Ray}\UFdesc
\author{J.~Raybern}\Dukedesc
\author{O.~Razuvaeva}\Mephidesc\ITEPnewadesc
\author{D.~Reyna}\Sandiadesc
\author{G.C.~Rich}\TUNLdesc
\author{J.~Ross}\NCCUdesc\TUNLdesc
\author{D.~Rudik}\Mephidesc
\author{J.~Runge}\Dukedesc\TUNLdesc
\author{D.J.~Salvat}\IUdesc
\author{A.M.~Salyapongse}\CMUdesc
\author{J.~Sander}\USDdesc
\author{K.~Scholberg}\Dukedesc
\author{A.~Shakirov}\Mephidesc
\author{G.~Simakov}\Mephidesc\ITEPnewadesc
\author{G.~Sinev}\altaffiliation{Also at: South Dakota School of Mines and Technology, Rapid City, SD, 57701, USA}\Dukedesc
\author{W.M.~Snow}\IUdesc
\author{V.~Sosnovtsev}\Mephidesc
\author{T.~Subedi}\VTdesc\Concorddesc
\author{B.~Suh}\IUdesc
\author{R.~Tayloe}\IUdesc
\author{K.~Tellez-Giron-Flores}\VTdesc
\author{E.~Ujah}\NCCUdesc\TUNLdesc
\author{J.~Vanderwerp}\IUdesc
\author{E.E.~van Nieuwenhuizen}\Dukedesc\TUNLdesc
\author{R.L.~Varner}\ORNLdesc
\author{C.J.~Virtue}\Laurentiandesc
\author{G.~Visser}\IUdesc
\author{K.~Walkup}\VTdesc
\author{E.M.~Ward}\UTKdesc
\author{T.~Wongjirad}\Tuftsdesc
\author{J.~Yoo}\SNUdesc
\author{C.-H.~Yu}\ORNLdesc
\author{J.~Zettlemoyer}\altaffiliation{Also at: Fermi National Accelerator Laboratory, Batavia, IL, 60510, USA}\IUdesc

\begin{abstract}
Using neutrinos produced at the Spallation Neutron Source (SNS) at Oak Ridge National Laboratory (ORNL), the COHERENT collaboration has studied the Pb($\nu_e$,X$n$) process with a lead neutrino-induced-neutron (NIN) detector. Data from this detector are fit jointly with previously collected COHERENT data on this process. A combined analysis of the two datasets yields a cross section that is $0.29^{+0.17}_{-0.16}$ times that predicted by the MARLEY event generator using experimentally-measured Gamow-Teller strength distributions, consistent with no NIN events at 1.8~$\sigma$. This is the first inelastic neutrino-nucleus process COHERENT has studied, among several planned exploiting the high flux of low-energy neutrinos produced at the SNS. 
\end{abstract}

\maketitle
%\tableofcontents

\section{Introduction}
\begin{figure*}[ht]
  \includegraphics[width=0.94\textwidth]{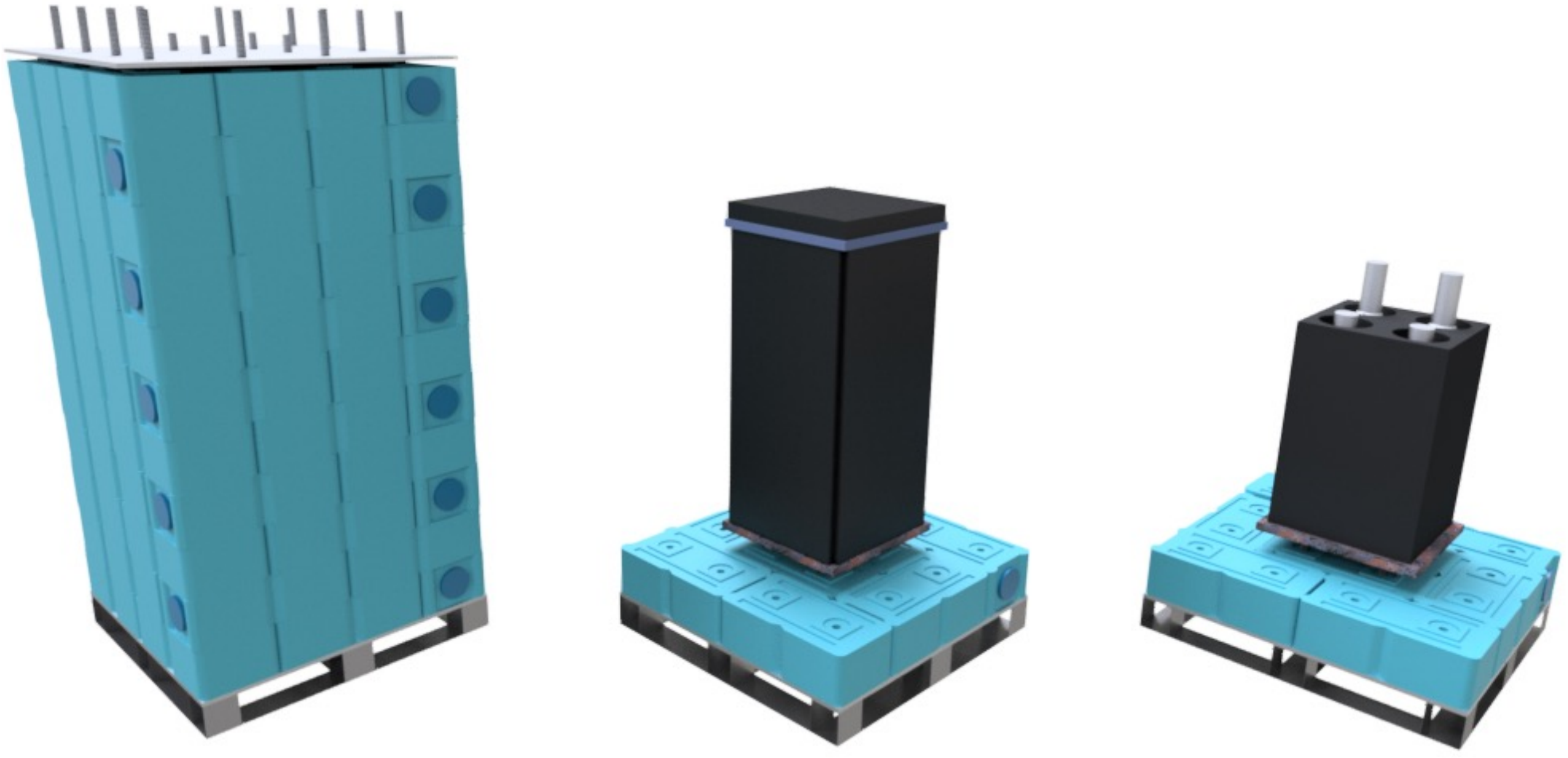}
  \caption{Left: exterior view of the neutrino cube showing water bricks used to reduce neutron backgrounds. Center: cut-away view with water bricks removed showing the muon veto panels. Right: cut-away view with muon veto panels removed showing the lead target along with two cylindrical and two hexagonal LS detectors. The bottom of the LS detectors is located approximately at the midpoint of the lead target.}
  \label{schematic}
\end{figure*}

When a sufficiently energetic neutrino interacts with a nucleus, neutrons can be emitted as part of the nuclear deexcitation process. Neutrino-induced neutrons (NINs) from low-energy ($\lesssim$\,50\,MeV) neutrinos  have been predicted since at least 1978~\cite{domogatskii1978}, but have not yet been experimentally observed. There are limited experimental measurements of low-energy neutrino-nucleus interactions through any channel~\cite{formaggio2012}, so cross section predictions for many interactions are untested. NINs produced through electron-neutrino charged-current (CC) interactions on lead are of particular interest, owing to the large predicted cross section~\cite{kolbe2001,volpe2002,suzuki2003,mclaughlin2004,athar2006,lazauskas2007,paar2008} and prevalence of lead as a shielding material. The electron neutrino CC reaction producing NINs is depicted in Eq.~(\ref{nineqn}) for ${}^{208}$Pb, the dominant isotope in naturally abundant lead. Multiple neutrons and/or gammas can be emitted as a result of the neutrino interaction.
\begin{equation}
\begin{aligned}
\nu_e + {}^{208}\mbox{Pb} \rightarrow e^- +& {}^{208}\mbox{Bi}^*\\
&{}^{208}\mbox{Bi}^*\rightarrow {}^{208-\mbox{x}}\mbox{Bi} + \mbox{X}n + \mbox{Y}\gamma
\end{aligned}
\label{nineqn}
\end{equation}
Neutrino-nucleus interactions can be used to detect supernovae neutrinos, impact supernova nucleosynthesis, and form backgrounds for neutrino and dark matter experiments.

The majority of existing supernova neutrino detectors are primarily sensitive to the electron-antineutrino component of the supernova neutrino flux through detection of inverse beta decay on hydrogen~\cite{rosso2021}. One of the exceptions is HALO~\cite{ZUBER2015233,duba2008}, which will detect supernova electron neutrinos through the production of NINs on lead. This provides HALO with a unique capability~\cite{GalloRosso2021}, although the NIN cross section on lead must be measured to determine detector sensitivity. NINs may also impact the observed nuclear recoil distribution in dark matter detectors with sensitivity to CEvNS interactions from supernova neutrinos~\cite{Bhattacharjee2022}.

Neutrino-nucleus reactions have been hypothesized to play a role in supernova nucleosynthesis~\cite{domogatskii1978}. This can occur through direct neutrino interactions ($\nu$-process) or through the production of NINs and subsequent rapid neutron capture ($r$-process)~\cite{domogatskii1978,woosley1990,sieverding2018,mclaughlin1996,qian1997}. A measurement of the NIN cross section for neutrinos of similar energy to those emitted by a supernova would help determine the impact of NINs on isotopic abundances generated in supernovae.

NIN interactions in detector shielding can potentially impact neutrino and dark matter experiments. In oscillation experiments, such as LSND~\cite{mills1998}, NINs produced in shielding may form a background~\cite{kolbe2001}. NINs from solar neutrino interactions have been proposed to explain the DAMA/LIBRA excess~\cite{davis2014}, although this claim has been refuted~\cite{barbeau2014,bernabei2014}. In searches for coherent elastic neutrino-nucleus scattering (CEvNS), NINs originating in shielding can produce low energy nuclear recoils in detectors, mimicking the CEvNS signal~\cite{akimov2017}. In particular, for CEvNS searches at pion decay-at-rest ($\pi$-DAR) neutrino sources~\cite{akimov2017,baxter202a}, NINs can be problematic as they follow the timing distribution of the CEvNS signal.

The COHERENT collaboration has studied the CC NIN process on lead using neutrinos produced at the Spallation Neutron Source (SNS) at Oak Ridge National Laboratory (ORNL)~\cite{PhysRevD.106.032003}. As an initial test, the collaboration deployed two 1.5-L liquid scintillator (LS) detectors inside shielding used for the COHERENT CsI[Na] detector~\cite{akimov2017} -- this detector is referred to as the ``Eljen cell" detector. The initial test observed a cross section $\sim$1.7 times lower than predicted in Ref.~\cite{mclaughlin2004}, although there were large uncertainties on the measurement. As a follow-up, two detectors -- referred to as the ``neutrino cubes"--were deployed to the SNS in 2015 to measure NIN production on lead and iron. This paper discusses the design of the lead neutrino cube, analysis of its $\sim$5-year exposure, and a combined result with an updated analysis of the Eljen cell detector.

%%%%%%%%%%%%%%%%%%%%%%%%%%%%
%%Experimental Description%%
%%%%%%%%%%%%%%%%%%%%%%%%%%%%
\section{Experimental Description}
\subsection{Detector overview}\label{sect:experimentalDescription}
The lead neutrino cube consisted of a $\sim$900-kg cast lead target with four cylindrical cavities for LS detectors. Two types of LS detectors were used during data collection: cylindrical 2.4-liter detectors and hexagonal 1.25-liter detectors. Both types held EJ-301 scintillator and used ET 9821-KEB 3'' photo multiplier tubes (PMTs). Plastic scintillator muon veto panels were placed against the sides and top of the lead target to reject muon-induced neutrons. Near-hermetic water shielding surrounded the veto panels to reduce environmental and beam-related neutrons (BRNs). A rendering of the detector can be found in Fig.~\ref{schematic}. For scale, the detector sits on a standard 36" $\times$ 36" $\times$ 6" pallet.

The detector was located 18.9\,m from the SNS target in ``Neutrino Alley'', where it was exposed to an intense flux of low energy ($<$52.8 MeV) neutrinos. At the SNS, bunches of protons strike a mercury target at a repetition rate of 60-Hz producing neutrons, $\pi^-$, and $\pi^+$. The majority of $\pi^-$ capture on mercury nuclei, while the $\pi^+$ come to rest before decaying (26\,ns lifetime) to produce prompt muon neutrinos and anti-muons. The anti-muons decay (2.2\,$\mu$s lifetime) to produce delayed muon-antineutrinos and electron neutrinos. As the full width at half maximum of the beam timing profile is $\sim350$~ns, prompt and delayed neutrino fluxes are separated in time, allowing isolation of $\nu_e$~CC NIN events from prompt events. More details on Neutrino Alley and neutrino production at the SNS can be found in Ref.~\cite{akimov2017}. 

The data analyzed were collected between 2016 and 2021. The detector operated in three configurations, (i) using four cylindrical detectors, (ii) using two cylindrical detectors (detectors further from target removed), and (iii) using two cylindrical and two hexagonal detectors (hexagonal detectors in slots further from target). The detectors were monitored for stable operation and periods with electronics issues or atypical proton beam were removed from analysis. Combining data from all configurations, the lead neutrino cube collected NINs over an exposure of 127\,GWHr$\cdot$liter, compared to the 10.05\,GWHr$\cdot$liter Eljen cell detector exposure. Neutrino production is parametrized as a function of proton beam energy and SNS target material in Ref.~\cite{PhysRevD.106.032003} with a 10\% normalization uncertainty that we have adopted in this analysis.

\subsection{Data collection and waveform reconstruction}
The lead neutrino cube triggered on the coincidence of a signal from any LS cell above a level threshold and a timing signal generated by the SNS within a $\sim$20\,$\mu$s window. The timing signal was synchronized to the proton pulse and used to determine the neutrino generation time. When a coincidence of these signals was detected, waveforms were recorded from all channels (LS cells, veto panels, and SNS timing signals).

A conditional moving average filter~\cite{grayson2017} was used to remove long-timescale oscillations in waveform baselines. To determine the start of a pulse, an interpolation algorithm~\cite{warburton2017} was applied to achieve a pulse onset time ($t_0$) with sub-sample precision, where onset was defined as the time at which the scintillation signal reaches 20\% of its maximum pulse amplitude. A 400\,ns window was integrated around the pulse onset (10\,ns prior, 390\,ns after) to determine the pulse energy. A pulse-shape discrimination (PSD) parameter was calculated as the ratio of the tail integral to the full integral of the pulse. The tail integral length was optimized for good separation of gammas and neutrons in each channel using a ${}^{252}$Cf neutron source (Sec.~\ref{neutronCalibration}) and varied from 355 to 360\,ns. The maximum height of each scintillation pulse is also recorded.

Cosmic events passing through the muon veto were identified by requiring a coincident signal in two or more muon veto PMTs within a 200\,ns window. Events in an LS cell were identified as muon-correlated if they occurred up to 200\,ns prior or up to 25\,$\mu$s after a muon event.

Events preceding the SNS timing signal by up to 2\,$\mu$s or following by up to 12\,$\mu$s were blinded during analysis to avoid developing cuts that bias the analysis.

\subsection{Energy calibration}
Calibrations were performed with dedicated gamma source runs several times throughout the detector's operational period, to determine the energy scale and energy resolution parameters of each LS cell. Gamma sources were simulated in MCNPX-PoliMI~\cite{mcnp}, added to background data collected during the calibrations, and fit to data, allowing the conversion from ADC-to-keV and energy resolution parameters to float. When data were collected with multiple gamma sources on the same date, the multiple datasets were fit simultaneously for each detector.

\begin{figure*}[ht!]
  \includegraphics[width=0.94\textwidth]{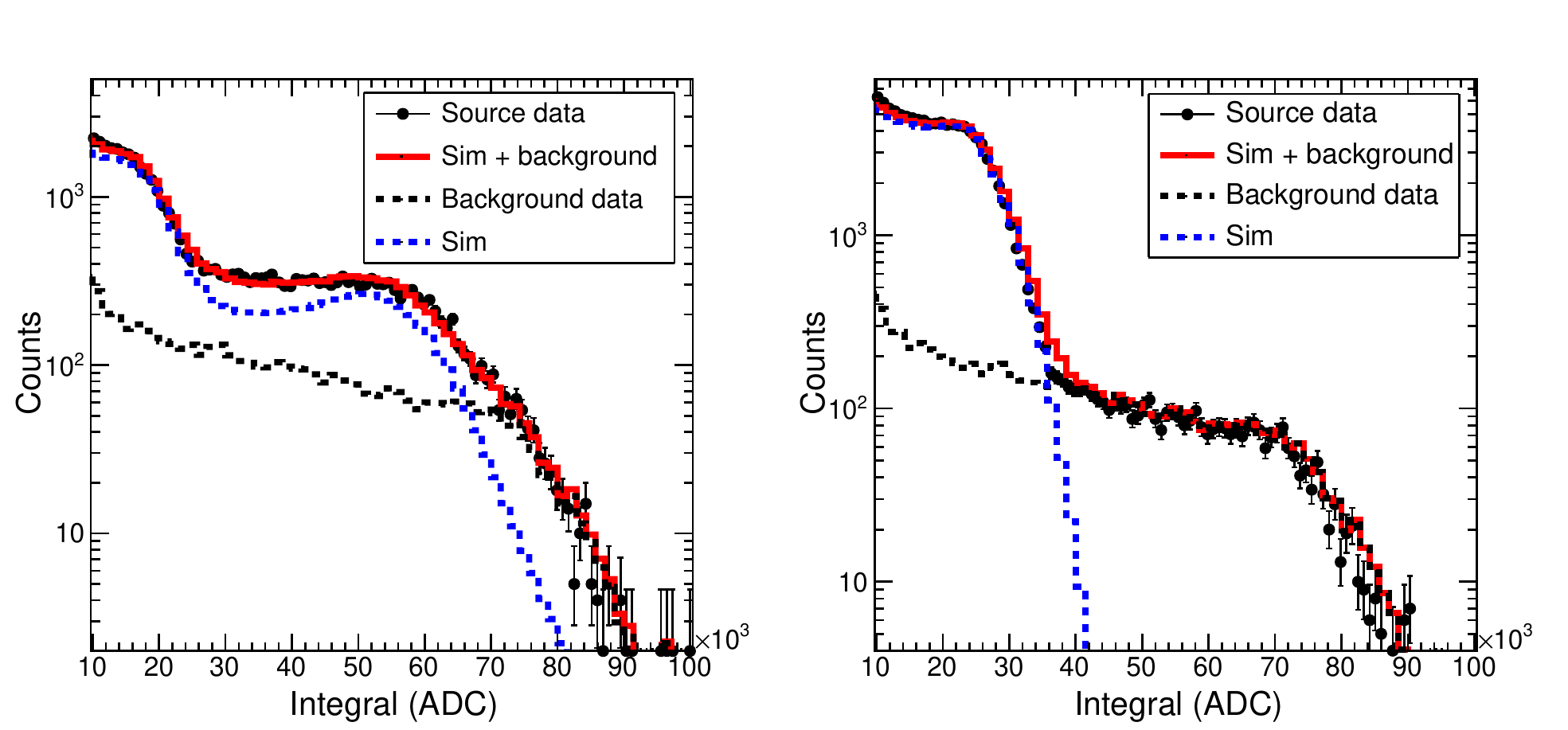}
  \caption{Using multiple gamma sources, the energy resolution and calibration parameters are determined from a simultaneous fit. A comparison is shown between data collected with a ${}^{22}$Na  (left) and  ${}^{137}$Cs source (right) along with simulated data after calibration and energy resolution is applied.}
  \label{sampleCalibration}
\end{figure*}

The light output in liquid scintillators for electronic interactions is approximately linear above $\sim$40 keV~\cite{dietze1982}. In that reference, the relationship between light output and true energy is described in Eq.~(\ref{calibrationEquation}):
\begin{equation}
	L = c(E_e-E_0).
	\label{calibrationEquation}
\end{equation}
Here $L$ is the light output in ADC, $c$ is a proportionality constant in units of keV-per-ADC, $E_e$ is the true energy in keV, and $E_0$ is a small offset accounting for nonlinearity of the light output at low energies~\cite{dietze1982}. $E_0$ was fixed to a value of 5 keV~\cite{byrd1992,arneodo1998,schmidt2002,klein2002}.

Energy resolution impacts the linear calibration parameter, and must be included in a fit of simulation to data to accurately determine the energy scale of the detectors. The energy resolution of a LS detector at an energy $E$ can be parametrized as
\begin{equation}
	\frac{\Delta E}{E} = \sqrt{\alpha^2+ \frac{\beta^2}{E} + \frac{\gamma^2}{E^2}},
	\label{energyresolution}
\end{equation}
where $\Delta E$ is the FWHM of a Gaussian centered at E. The parameters ($\alpha$,$\beta$,$\gamma$) originate from different aspects of the light production~\cite{dietze1982}. A sample calibration can be found in Fig.~\ref{sampleCalibration}. Uncertainties in the energy calibration are incorporated into our systematic uncertainties in Table~\ref{table:Syst}. By combining log-likehood distributions, a single set of energy resolution parameters is obtained for each detector from all gamma calibrations over the detector's exposure -- the resulting systematic uncertainty on energy resolution is accounted for in Table~\ref{table:Syst}.

The dedicated gamma source calibrations allow a conversion from ADC-to-keV to be known at the time of the calibration. However, this conversion factor was observed to change over time. Potential sources of this drift could be PMT aging, changes in temperature, oxygen leaking into the scintillator, or helium leaking into the PMTs~\cite{photonics2000photomultiplier}. To correct for time-dependent gain drift, a procedure was developed to fit the high energy background spectrum from each run to a spectrum obtained at the start of the data collection period, allowing for the determination of the relative gain drift of the detectors over time. The high energy region features a strong signal from ${}^{40}$K, likely originating from the phototube. At lower energies, the background spectrum shape may change over time due to time-varying external backgrounds present in Neutrino Alley (predominantly 511-keV $\gamma$-rays).

A RooKeysPdf~\cite{Verkerke:2003ir} was generated from the first ninety-six hours of operation of each liquid scintillator cell. This RooKeysPdf was fit to every subsequent ninety-six hour period of data collection throughout the detector's operation, allowing a single scaling factor to float. Following the fit, a spline was formed to interpolate the gain correction factor for individual runs within this ninety-six hour period. The relative gain curves were fit to the known absolute ADC-to-keV conversions determined by the dedicated gamma source calibrations.

\subsection{Neutron calibration}\label{neutronCalibration}

\begin{figure}[!ht]
\centering
\vspace*{1ex}\includegraphics[width=8.6cm]{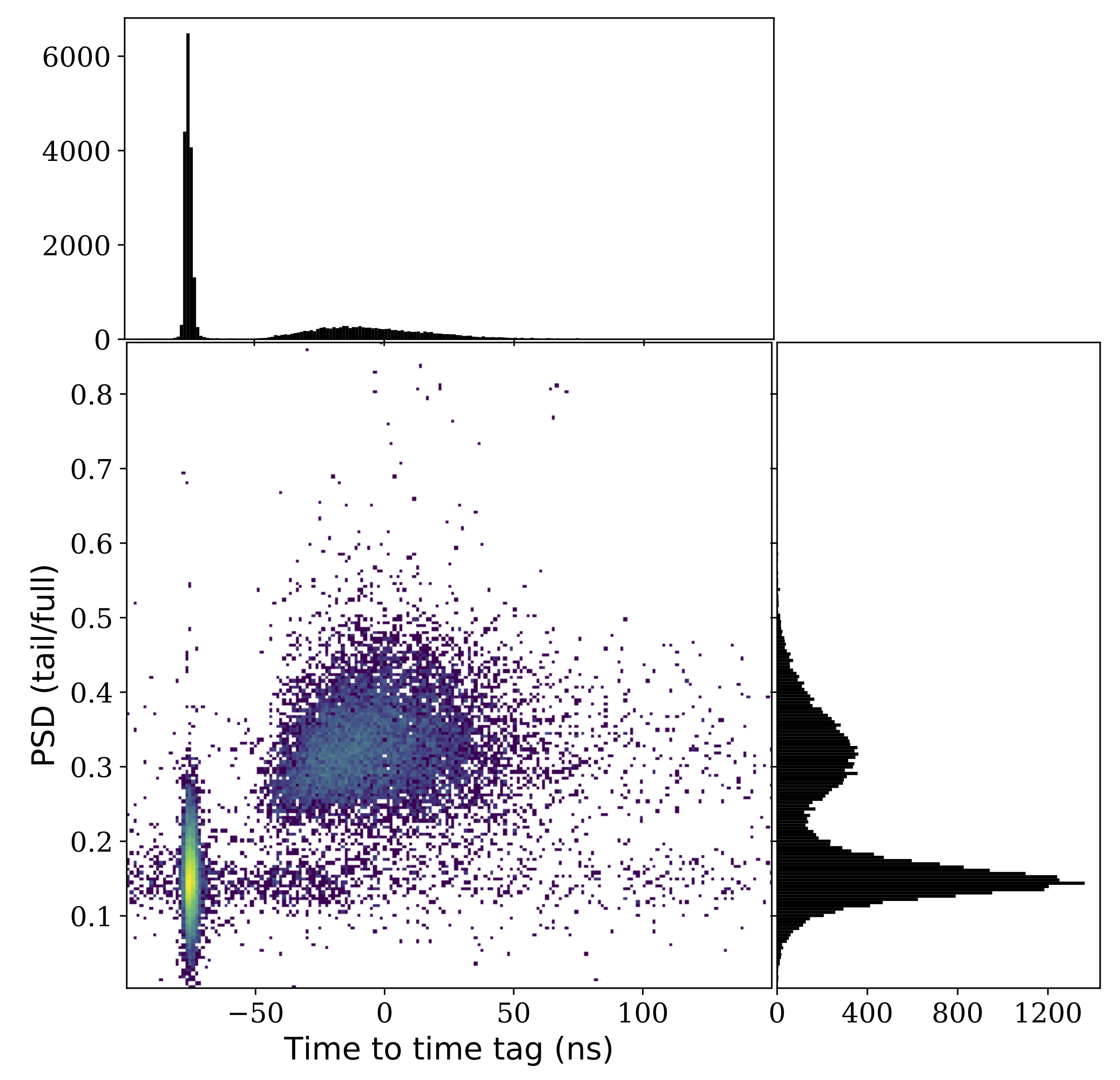}
\caption{Time of events in a liquid scintillator cell relative to the time-tagged decay compared to the PSD distribution. Gamma rays occur at approximately t=-80\,ns, and neutrons occur between -50 to 200\,ns.}
\label{timeTaggedFigure}
\end{figure}

A time-tagged ${}^{252}$Cf source (produced at the Radiochemical Engineering Development Center (REDC) at ORNL) was used to identify a clean population of neutrons for optimizing PSD parameters and determining trigger efficiencies. The time-tagged source consists of ${}^{252}$Cf inside a small biased ionization chamber. The 
${}^{252}$Cf spontaneous fission can produce fission fragments, which are detected in the ionization chamber as a signature that a fission occurred, along with gammas and neutrons. For these runs, the time-tagged signal replaced the SNS timing signal in the data acquisition system, but otherwise the detector data acquisition configuration remained unchanged. The PSD and timing distribution from this calibration are shown in Fig.~\ref{timeTaggedFigure}.

\begin{figure}[!bt]
\centering
\includegraphics[width=8.6cm]{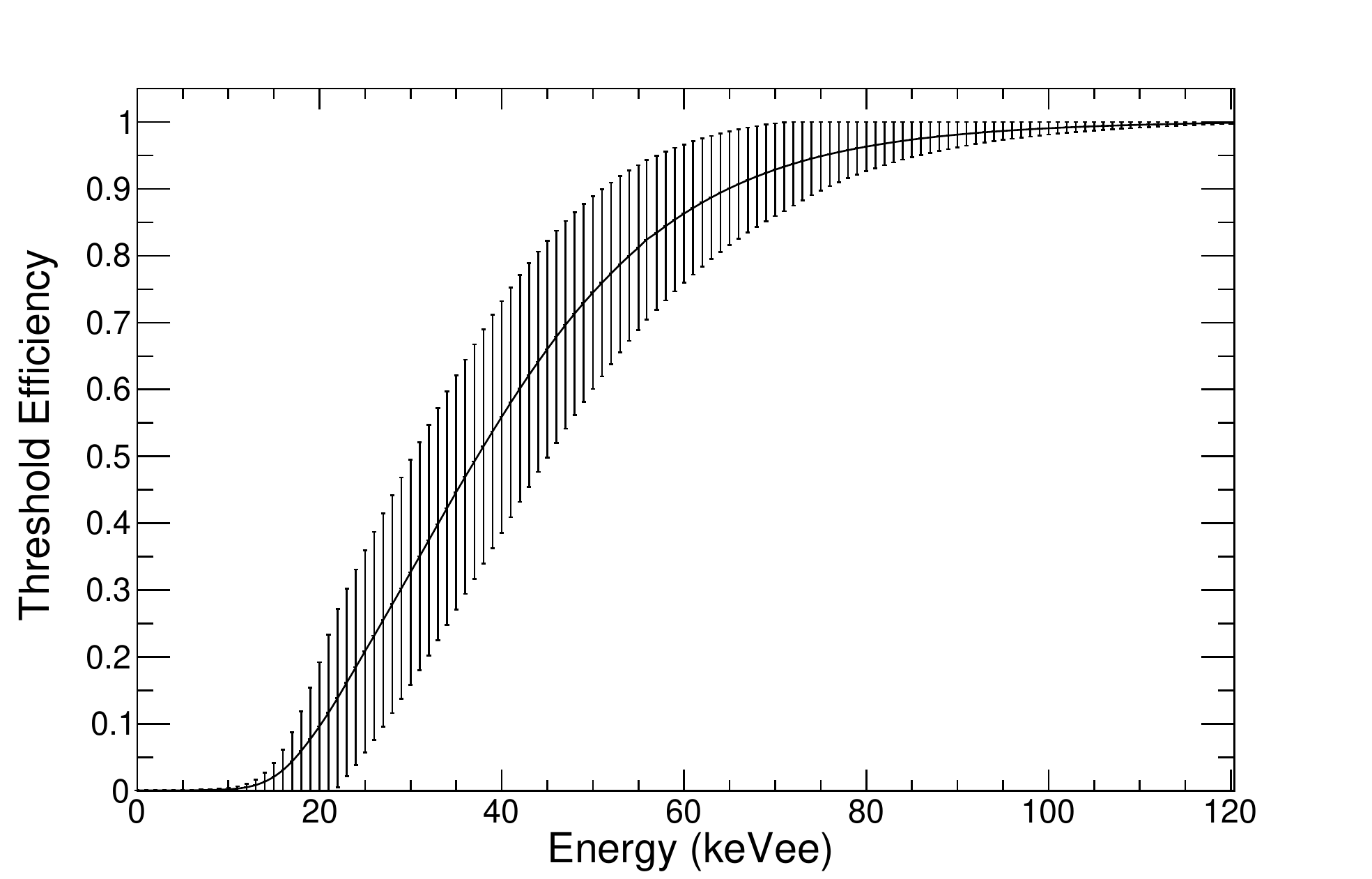}
\caption{Efficiency of the software trigger to neutron pulses from all LS cells, weighted by each cell's exposure. Errors are derived from uncertainties fitting parameters.}
\label{thresholds}
\end{figure}
Using the populations of gamma rays and neutrons identified with timing from the ${}^{252}$Cf calibration, software thresholds were determined for gamma rays and neutrons. This was done by quantifying the relationship between pulse height and integral using data, and fitting the peak height distribution associated with events of a known integral. Because gamma rays and neutrons have different pulse shapes, their trigger efficiencies will be different, and thus they were determined independently. By incorporating individual detector gain change and beam exposure, a single weighted trigger efficiency curve was produced as a function of nuclear recoil energy, shown in Fig.~\ref{thresholds}.

\begin{figure}[!bt]
\centering
\includegraphics[width=8.6cm]{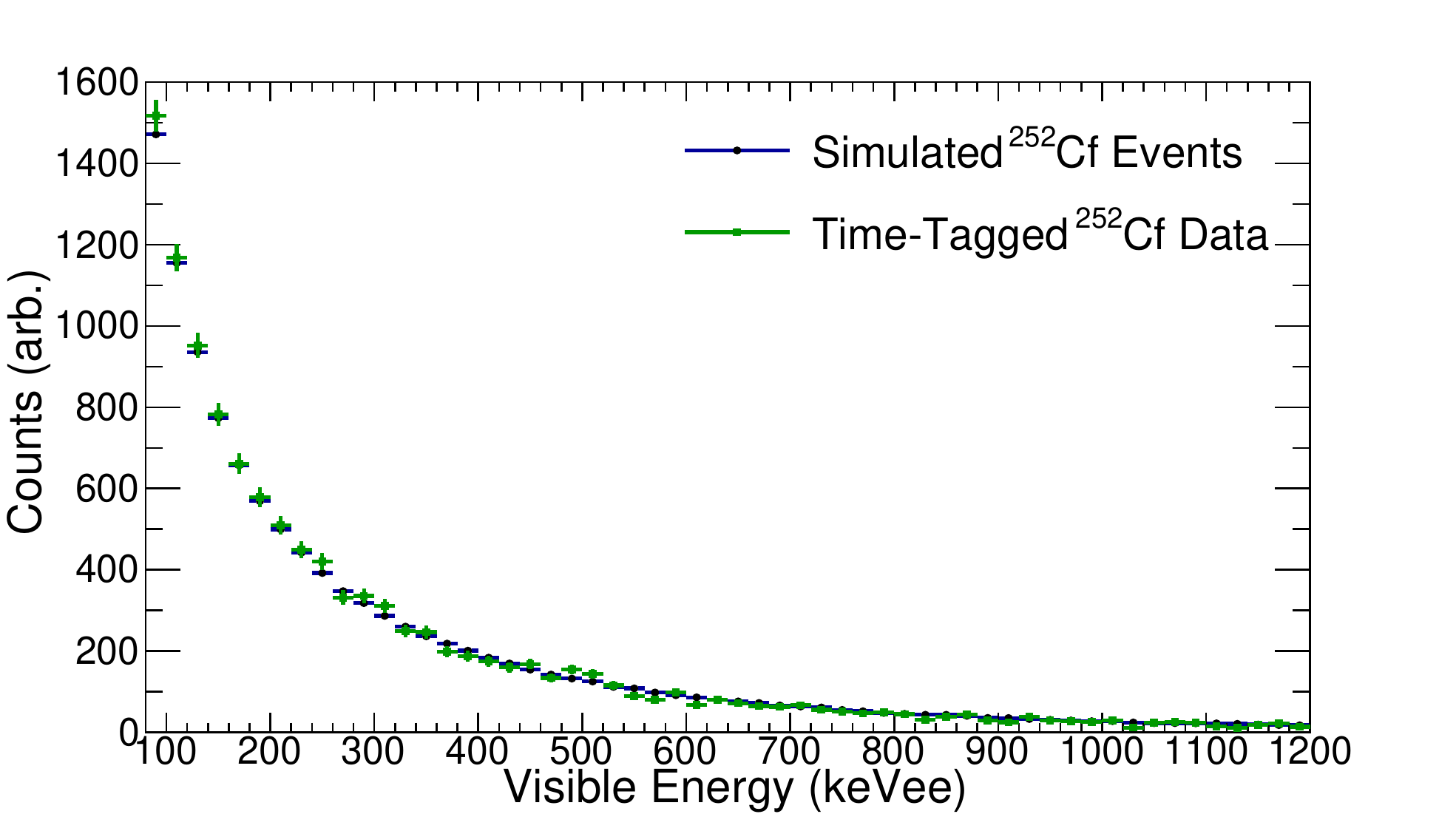}
\caption{Comparison of simulated ${}^{252}$Cf events and data for a single channel in the energy region-of-interest. The simulated data have energy resolution, trigger thresholds, and nuclear recoil quenching factors applied.}
\label{timeTaggedFigure_2}
\end{figure}

As a check, the observed spectrum of neutrons from the time-tagged ${}^{252}$Cf source is plotted along with simulation in the NIN energy region of interest in Fig.~\ref{timeTaggedFigure_2}. In both simulation and data, events are selected within a 140\,ns window encapsulating the neutrons originating from the source. The simulation has energy resolution, trigger thresholds, and nuclear recoil quenching factors from EJ-301~\cite{awe2018,verbinski1968} applied, and shows excellent agreement with the data.

\subsection{Pulse-shape discrimination}\label{sect:PSD}
\begin{figure*}[ht!]
  \includegraphics[width=\textwidth]{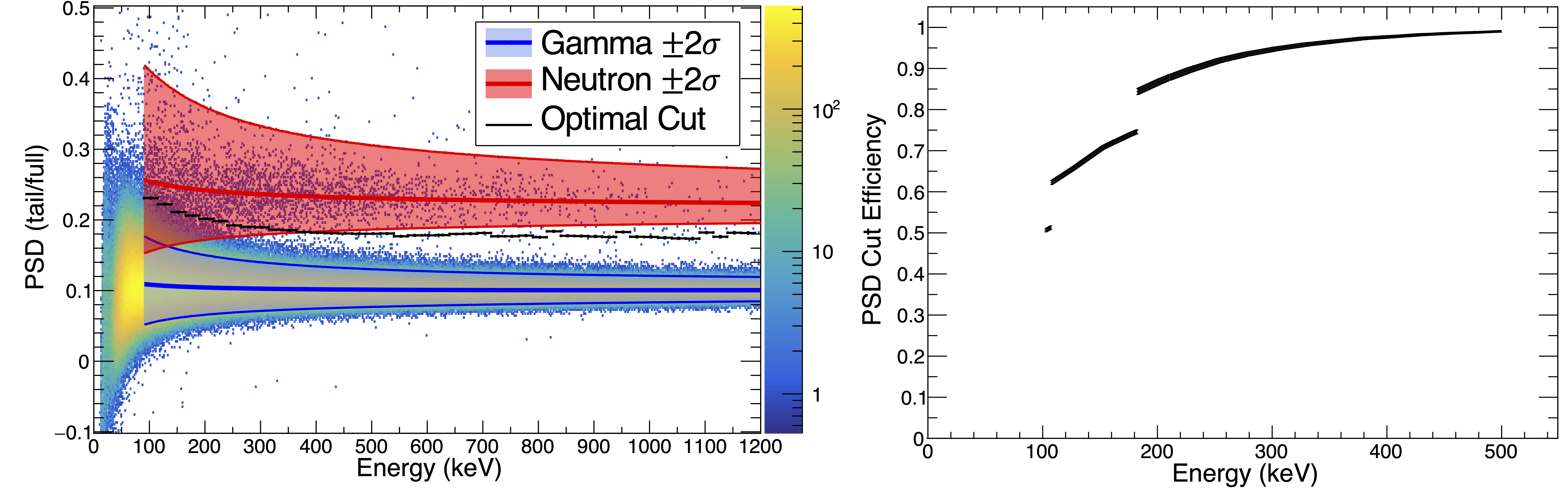}
  \caption{Left: PSD distribution for ${}^{252}$Cf events tagged in a single LS detector, with the 2$\sigma$ acceptance region shown in red for neutrons and blue for gamma rays. The black line shows the optimal PSD cut used in the analysis for this channel. Right: The optimized PSD cut for each channel is weighted by that channel's exposure to produce a PSD efficiency curve. The discontinuities arise from variations in the minimum PSD threshold achieved for each LS.}
  \label{psdFigures}
\end{figure*}

The PSD distribution of events in LS cells were observed to vary with time -- this can result from PMT aging, oxygen leaking into the LS, or helium leaking into PMTs~\cite{photonics2000photomultiplier}. By studying PSD distributions over long periods of time, gamma and neutron PSD distributions were observed to drift by the same factors.  The variation of PSD was corrected by tracking the means of the gamma PSD distributions and scaling the PSD parameter such that the gamma mean always occurred at the same value. The distributions of corrected PSD values from blinded data were fit using the convolution of a Gaussian and an exponential decay to model the shapes of the neutron and gamma populations. The energy-dependence of the free parameters were derived from the ${}^{252}$Cf calibration, but were allowed to float in the fit to blinded data. A cut in PSD space was imposed to maximize the sensitivity of each LS cell using predicted signal rates (Sec.~\ref{signalPrediction}). The lowest energy for which good separation between gamma rays and neutrons could be achieved varied for each detector, ranging from 90 to 170\,keV. PSD threshold curves from each detector were weighted by beam exposure to produce a single PSD efficiency function. The variation in lower PSD thresholds for different channels leads to discontinuities in the weighted efficiency of the entire detector. A typical PSD distribution, along with the weighted PSD cut efficiency, is depicted in Fig.~\ref{psdFigures}. 

%%%%%%%%%%%%%%%%%%%%%
%%Signal Prediction%%
%%%%%%%%%%%%%%%%%%%%%
\section{Signal Prediction}\label{signalPrediction}
Predictions for CC events were obtained using the MARLEY event generator~\cite{marley,gardiner2021}. While MARLEY was originally developed for modeling CC interactions on argon~\cite{gardiner2021_2}, it can be adapted for use with other nuclei by supplying the corresponding Gamow-Teller (GT${}^-$) and Fermi (F) strength distributions. For ${}^{208}$Pb, GT${}^-$ strengths were obtained from the data in Ref.~\cite{wakasa2012}, provided by the authors for our use. 

MARLEY simulates neutrino-nucleus interactions using the allowed approximation (neglecting Fermi motion and in the zero-momentum transfer limit). This is an approximation for $\pi$-DAR neutrinos, as forbidden transitions begin to play a more important role at higher energies~\cite{volpe2002}. While MARLEY’s treatment of CC interactions is incomplete, it is currently the only event generator available for simulating CC neutrino-nucleus events from stopped-pion neutrinos for a variety of nuclei.

The experimentally-measured GT${}^-$ strengths were scaled by $g_A^2=(1.26)^2$ to account for the different definition of matrix elements in charge-exchange and weak interactions--this value was chosen as it is the value adopted in Ref.~\cite{taddeucci1987} which is the source of the normalization of the Gamow-Teller strength in charge-exchange reactions. Additionally, the energy scale was adjusted to account for the difference in the ground state energy of ${}^{208}$Pb and ${}^{208}$Bi. The Fermi strength was obtained using the Fermi sum rule, $B(F)=N-Z$, with an energy centered on the isobaric analog state of ${}^{208}$Pb in ${}^{208}$Bi~\cite{osterfeld1992}. The predicted inclusive MARLEY flux-averaged cross section for $\pi$-DAR electron neutrino CC interactions with ${}^{208}$Pb is $42.1 \times 10^{-40}\mbox{cm}^2$, in good agreement with existing predictions (ranging from $\sim26-50 \times 10^{-40}\mbox{cm}^2$) ~\cite{kolbe2001,volpe2002,suzuki2003,mclaughlin2004,athar2006,lazauskas2007,paar2008}. 

As in Ref.~\cite{kolbe2001}, a crude $N-Z$ scaling of the cross section was assumed for the naturally occurring isotopes of lead in the signal calculation. This was done because there are no experimental measures of the Gamow-Teller strength for other naturally occurring isotopes of lead. The inclusive cross section from MARLEY for ${}^{208}$Pb, along with the partial cross sections leading to neutron emission, are shown in Table~\ref{marleyTable}. 

{\renewcommand{\arraystretch}{1.3}
\begin{table}[!tb]
\begin{center}
\begin{tabular}{cc}
\hline\hline
Channel & Cross section ($\times 10^{-40}\mbox{cm}^2)$\\
\hline
${}^{208}\mbox{Pb}(\nu_e,X)$ & $42.1_{-4.7}^{+4.7}$\\
${}^{208}\mbox{Pb}(\nu_e,e^{-}+n){}^{207}\mbox{Bi}$ & $31.7_{-3.2}^{+3.1}$\\
${}^{208}\mbox{Pb}(\nu_e,e^{-}+2n){}^{206}\mbox{Bi}$ & $7.6_{-1.4}^{+1.5}$\\
${}^{208}\mbox{Pb}(\nu_e,e^{-}+3n){}^{205}\mbox{Bi}$ & $0.4_{-0.1}^{+0.0}$\\
\hline\hline
\end{tabular}
\caption{Predictions of CC cross sections for ${}^{208}$Pb using $\pi$-DAR electron neutrinos with MARLEY configured with B(GT${}^{-}$) data from Ref.~\cite{wakasa2012}. Interaction channels which do not produce a final-state neutron are included in the $^{208}$Pb($\nu_e$,$X$) value. For lead of naturally-occurring isotopic abundances, imposing $N-Z$ scaling produces an inclusive flux-averaged cross section of $41.4 \times 10^{-40} \mbox{cm}^2$. All uncertainties originate from those on the measured GT matrix elements, and does not include additional uncertainties from the inclusion of forbidden transitions or on the MARLEY model.}
\label{marleyTable}
\end{center}
\end{table}

CC events generated by MARLEY were simulated using Geant4~\cite{geant} with the detector geometry to determine an average efficiency of NINs reaching the LS cells of 18.8\%. The simulation output was processed matching analysis cuts (EJ-301 quenching factors from Refs.~\cite{awe2018,verbinski1968}, fit energy resolution parameters, measured trigger thresholds, and measured PSD efficiencies). Applying these reduces the average NIN detection efficiency to 3.3\%. While MARLEY does not interface directly with MCNP, as a cross-check neutrons were simulated in 500-keV bins in MCNPX-PoliMi\cite{mcnp}. Although the NIN selection efficiency could only be approximately determined with the 500-keV bins implemented in the MCNP simulation, it agreed with the GEANT simulation to within 4.3\%.

MARLEY can also be used to simulate neutral-current (NC) interactions, although it can only simulate allowed transitions. To provide MARLEY with the $B(GT_0)$ matrix elements to generate NC predictions, the experimentally measured magnetic dipole transition strength distribution, $B(M1)$, from~\cite{poltoratska2012,birkhan2016} was used along with an assumed conversion from $B(M1)$ to $B(GT_0)$ from Ref.~\cite{semenov2017} (see Refs.~\cite{semenov2017,lee1978,langanke2004} for a discussion on the proportionality of $B(M1)$ strengths to $B(GT_0)$ strengths). This results in predicted flux-averaged inclusive NC cross sections of 0.7$\times 10^{-40}\mbox{cm}^2$, 0.5$\times 10^{-40}\mbox{cm}^2$, and 1.0$\times 10^{-40}\mbox{cm}^2$ for $\nu_e$, $\nu_{\mu}$, and $\bar{\nu}_{\mu}$ respectively. There are few existing calculations~\cite{suzuki2003,semenov2017} to compare with MARLEY's predictions for the inclusive NC cross section on lead for $\pi$-DAR neutrinos, and no other predictions for NC NIN channels from these sources. The predicted fraction of NC NIN events is small, as it only depends on the $B(M1)$ strength above the neutron emission threshold in ${}^{208}$Pb. This $\sim1.6\%$ contribution to the expected signal rate is not included in the fit, but is incorporated as a systematic uncertainty. As a comparison, the NC component expected in HALO is larger as all flavors of neutrinos are emitted by supernovae, while only $\nu_e$, $\nu_{\mu}$, and $\bar{\nu}_{\mu}$ are produced at the SNS in significant quantities.

Calculations in Ref.~\cite{PhysRevD.106.032003} were used to determine the expected number of neutrinos generated as a function of proton energy and SNS beam power. Using the nominal MARLEY cross section and simulated efficiencies, $346_{-59}^{+58}$ CC NIN events were expected in the analyzed dataset.

%%%%%%%%%%%%%%%%%%%%%%%%
%%Results & Discussion%%
%%%%%%%%%%%%%%%%%%%%%%%%
\section{Results \& Discussion}
\subsection{Lead neutrino cube}
The primary challenge in identifying NIN events above threshold is discriminating between NINs and prompt BRN backgrounds. Both produce proton recoils and thus have similar PSD distributions. The recoil energy distributions of the two populations are also similar. However, NINs and BRNs occur at different times. The BRNs are expected to closely follow the proton bunch timing distribution, whereas the electron neutrinos are delayed (see Sec.~\ref{sect:experimentalDescription}). Thus, a 1D fit in recoil time was performed to determine the NIN signal and prompt background counts. All counts that pass PSD selection cuts and have a reconstructed energy less than a detector-specific upper limit ranging from 425--825 keV$_{ee}$ were included in the fit. The upper energy limit for each detector was determined to optimize separation of NIN signal and steady-state alpha backgrounds, which can have PSD parameters similar to neutrons.

A number of sources of systematic uncertainty were evaluated that affect our determination of the NIN cross section. These are included in the result and listed in Table~\ref{table:Syst}. The dominant normalization uncertainty originates in the uncertainty in neutrino flux at the SNS, which will be improved with future COHERENT data~\cite{COHERENT:2021xhx}. Quenching and calibration uncertainties are the next largest sources, each affecting the cross section by $2-3\%$. The timing of the NIN pulse was determined from measurements of the SNS beam current with an uncertainty of $\sim$38\,ns. We also studied uncertainties in neutron scattering with nuclei, but determined these sources were negligible. There is no appreciable attenuation of the neutron flux between the NIN interaction point and the scintillator cells due to the poor stopping power of neutrons traveling through lead; thus, the result is insensitive to the $n-^{208}$Pb uncertainty.  Further, since C recoils are strongly quenched in scintillator, we must only consider $n-p$ interactions in the scintillator.  This interaction is well understood and has been measured at $<1\%$ for few-MeV neutrons~\cite{PhysRev.94.389} and is thus a negligible uncertainty.

\begin{table}[!tb]
\begin{center}
\begin{tabular}{cc}
\hline\hline
Source & NIN uncertainty ($\%$) \\
\hline
Neutrino flux & $\pm$10 \\
Quenching factor & $\pm$2.7 \\
Software threshold & $+0.2$ / $-0.4$ \\
PSD selection & $\pm$1.0 \\
Calibration & $+2.1$ / $-2.2$ \\
Energy resolution & $+1.7$ / $-0.5$ \\
Muon veto & $+0.4$ / $-0.3$ \\
Lead target mass & $\pm$0.6 \\
MARLEY NC prediction & $+0$/$-1.6$ \\
\hline
Total: & $+10.8$ / $-10.8$\\
\hline\hline
\end{tabular}
\caption{Summary of sources of systematic uncertainty which affect NIN normalization.}
\label{table:Syst}
\end{center}
\end{table}

The normalization of the prompt BRN background was allowed to float freely. The timing distribution of these events is critical for determining the NIN rate, so the mean and width of the neutron timing pulse were included as unconstrained parameters in the fit. Measured neutron time-of-flight depends on detector threshold, which governs the neutron energy range producing the background, and location within Neutrino Alley, so previous neutron data from COHERENT may not accurately constrain these uncertainties. The width of the BRN pulse may be wider than the protons-on-target (POT) pulse due to variations in neutron time-of-flight. This broadening is incorporated into the fit by convolving the POT trace with a Gaussian smearing of BRN arrival times. The standard deviation of the Gaussian is treated as a free parameter.

After determining reconstruction, selection, and analysis methods, the data were unblinded, resulting in the spectrum shown in Fig.~\ref{fig:BFdata}. The data selected by the cuts were fit to determine the NIN normalization. Two independently developed fitting codes evaluated the data to confirm consistency of the analysis procedure. An unbinned likelihood fit revealed $36^{+72}_{-36}$ NIN events in the sample, while a binned likelihood fit of the same data found $37^{+69}_{-37}$ NIN events allowed at 1$\sigma$, with both fits producing consistent results. The fit estimated $1,295_{-48}^{+44}$ BRNs with an additional $79^{+11}_{-11}$~ns of arrival time smearing. The inferred NIN rate is $>4\sigma$ lower than expectations from MARLEY.

\begin{figure}[!bt]
\centering
\includegraphics[width=8.6cm]{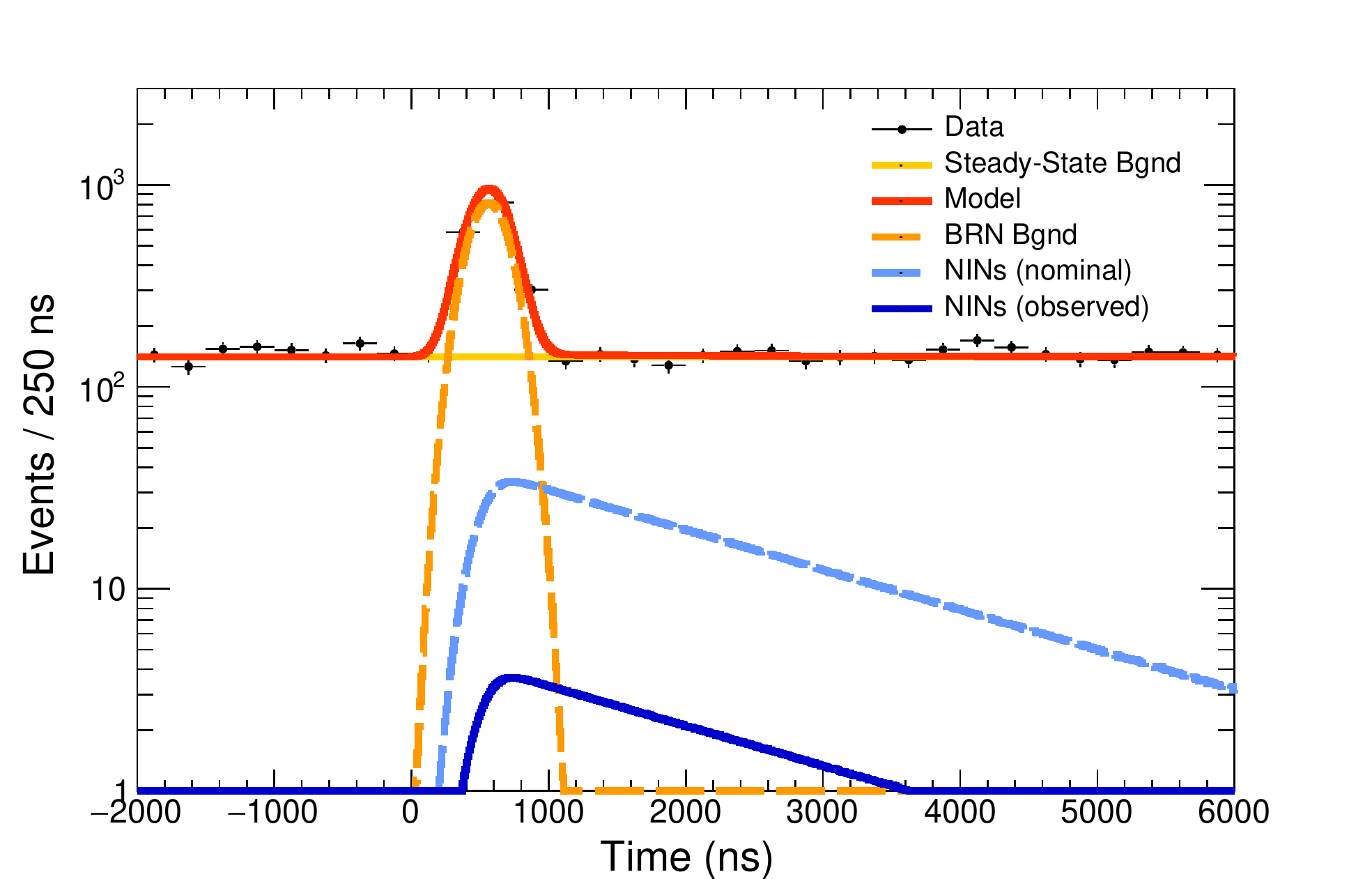}
\caption{Timing distribution of observed data within the SNS beam window along with detector background and the predicted and best-fit NIN contributions. The data are consistent with the no-NIN scenario, with a 1$\sigma$ range of $37^{+72}_{-37}$ NIN counts.}
\label{fig:BFdata}
\end{figure}

As this result is discrepant with expectations, several checks were subsequently performed. First, the lead used for the target was checked for impurities. The presence of lighter atoms in the lead may increase the neutron scattering cross section in the target and decrease the target density. The lead is stamped $99.99\%$ natural lead. The density of lead used to cast the target was measured to be $11.48\pm0.21$\,$\mbox{g/cm}^3$, within uncertainty of the nominal value of $11.29$\,$\mbox{g/cm}^3$. Second, the rate of neutrons in the prompt and delayed windows (applying all neutron selection cuts) were compared to the delivered beam exposure to check the stability of the neutron selection efficiency during detector operations. These distributions are plotted in Fig.~\ref{fig:ks}. Third, the PSD cut was extended to include lower energy recoils. This increases the expected number of detected neutrons and increases sensitivity to lower energy neutrons. Lowering the PSD cut to 50~keV increased the number of expected NINs by a factor of 1.43. However, these lower energy recoils were not included in the original fit as the uncertainty on their selection efficiency is large. Fitting this sideband gave a NIN rate consistent with that observed in the nominal analysis sample.

\begin{figure}[!ht]
\centering
\includegraphics[width=8.6cm]{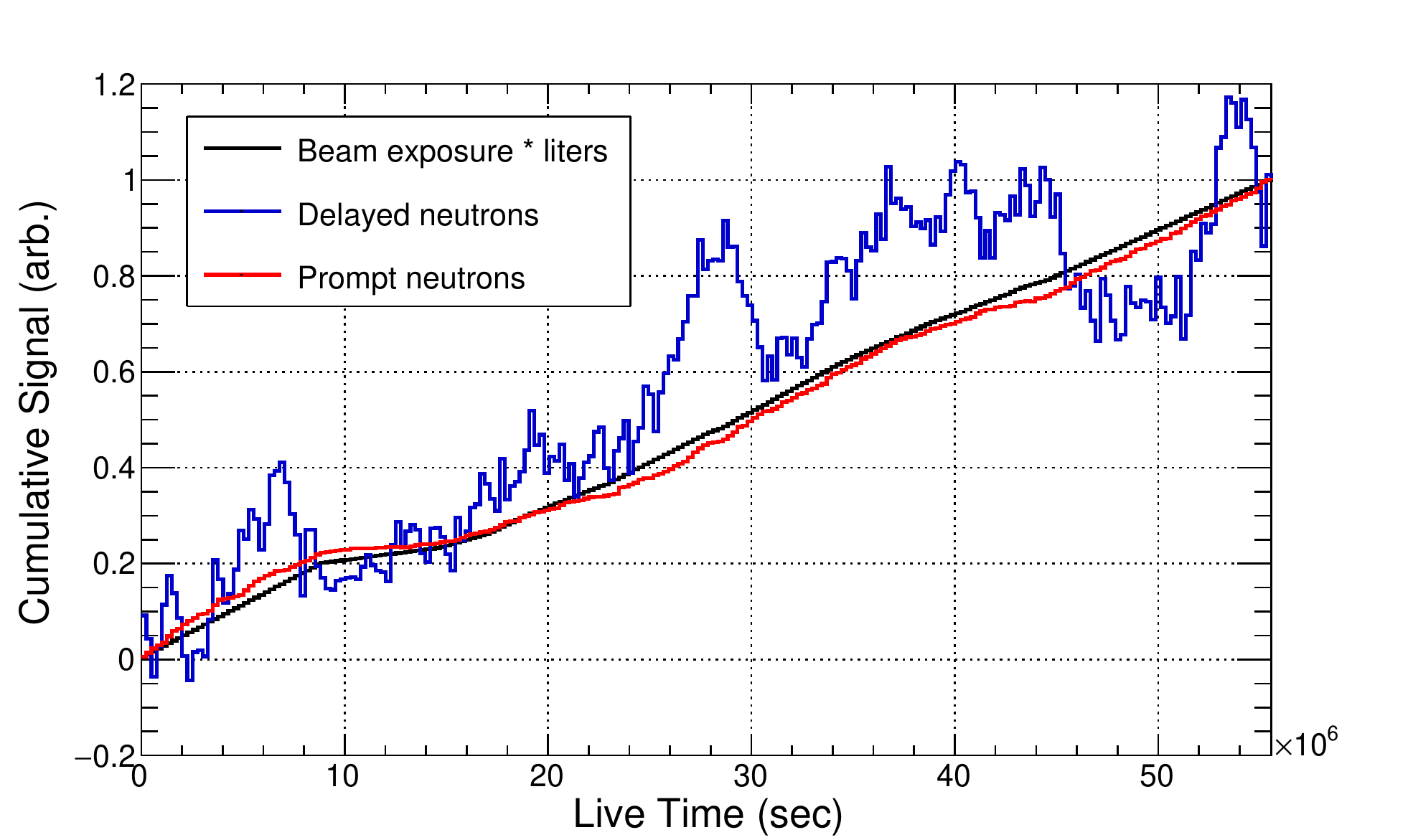}
\caption{Comparison of the cumulated beam exposure in the lead neutrino cube (scaled by active liquid scintillator liters for each run) to background-subtracted prompt and delayed neutron excesses.}
\label{fig:ks}
\end{figure}

\subsection{Combined results}
The analysis of the Eljen cell data in Ref.~\cite{akimov2017} has been updated. The previous analysis assumed the emitted NINs followed an evaporative neutron spectrum, and assumed the spectra of neutron events of all multiplicities were identical. The updated analysis used MARLEY to generate neutron energy distributions that depend on the neutron multiplicity of events along with new calculations of the number of neutrinos produced per incident POT from Ref.~\cite{PhysRevD.106.032003}. Finally, time broadening of the BRN population was incorporated into the fit and allowed to float, as is done in the analysis of the lead neutrino cube, to incorporate BRN time-of-flight dispersion effects. One main difference between the two datasets is the PSD thresholds, which are 30\,keV for the Eljen cell detector and 95--170\,keV for the lead neutrino cube detectors.

Combining the results yields a cross section scaling factor of  $0.29^{+0.17}_{-0.16}$ relative to predictions from MARLEY. For the neutrino cube sample, this corresponds to 100$^{+57}_{-54}$ events, consistent with the fit using only data from this detector. The likelihood profiles from the combined fit can be found in Fig.~\ref{fig:LL}.

\begin{figure}[!bt]
\centering
\includegraphics[width=8.6cm]{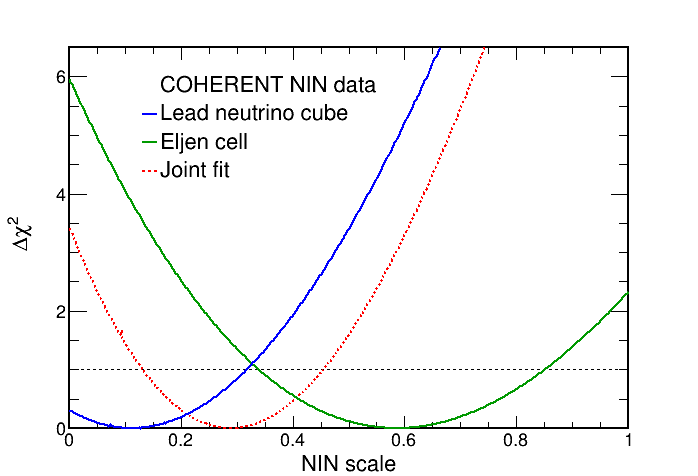}
\caption{Likelihood profiles from the lead neutrino cube and Eljen cell detector updated analysis. A value of 1 corresponds to the predicted signal from the MARLEY event generator.}
\label{fig:LL}
\end{figure} 

\subsection{Discussion}
The lead neutrino cube was designed to measure the ${}^{nat}\mbox{Pb}(\nu_e,e^{-}+\mbox{X}n)$ cross section from SNS electron neutrinos. Its measurement of the process, along with a combined fit with data from the Eljen cell detector, yield an observed reduction in signal compared to prediction. The nature and cause of this suppression is not known, but several possibilities are discussed below. The inclusion of forbidden transitions into the signal prediction would increase the theoretical cross section, increasing the observed reduction, although contributions from forbidden transitions have not been measured experimentally.

One explanation is the predicted inclusive CC cross section could be lower than expected, affecting neutron emitting and non-neutron emitting channels. A suppression in $g_A$ would reduce the inclusive cross section. This hypothesis could be tested with a dedicated detector measuring the electromagnetic energy deposited by CC events. This can be done within COHERENT, and there are also external measurement plans (see Ref.~\cite{pidar}). Additionally, data from a detector measuring the electromagnetic component of CC interactions on ${}^{127}$I have been collected for several years at the SNS. These results also show the measured cross section is lower than predicted by MARLEY~\cite{COHERENT:2023ffx}; these data will help test theoretical calculations of the neutrino cross section and final-state particles.

Another possibility is that neutrons emitted by CC interactions have lower energies than predicted by MARLEY. If so, the sensitivity of the HALO experiment would be largely unaffected by the observed reduction. There are no existing models predicting the emission of lower energy neutrons from CC events on lead, but it remains a possibility. A measurement of NIN production with a capture-gated detector would test this.

Measuring NINs on a lighter target may also clarify the current result. In 2017, the iron neutrino cube was deployed to the SNS, seeking to measure NINs produced on iron with a $\sim$700-kg target. If the NIN cross section on iron is similarly lower than predictions as current data suggest, it will be more difficult to observe NINs in the iron neutrino cube, but analysis is still underway on its data. Additionally, an analysis of the 2$n$ NIN emission cross section on lead is planned using data collected by the lead neutrino cube. While the rate of observed events is expected to be lower, backgrounds are also significantly reduced.

There are several implications of a reduced lead NIN cross section. For CEvNS detectors at spallation sources, this measurement reduces the expected backgrounds and impacts design of detector shielding. This result may have negative implications for HALO's potential to study supernova neutrinos, though the experiment would not be affected if the decreased observed NIN rate is a consequence of a softer neutron spectrum rather than a lower rate than predictions. However, HALO continues to be vital for understanding the next galactic core-collapse supernova with its sensitivity to the $\nu_e$ component of the supernova flux.

\section{Conclusion} Five years of data were analyzed to study NINs produced from electron neutrino CC interactions on lead at the SNS. Combining this result with an updated analysis of the Eljen cell detector yields a cross section suppressed by $0.29^{+0.17}_{-0.16}$ compared to the MARLEY prediction. The cause of the observed reduction is unknown, but future experiments will help to determine its origin. Within COHERENT, updated measurements of the neutrino flux with a heavy-water detector will improve systematic uncertainties on the existing measurements \cite{COHERENT:2021xhx}, and measurements of CC interactions on other targets may help determine whether a similar suppression is observed with other nuclei.

\section{Acknowledgments}
The COHERENT collaboration would like to thank the Duke Machine Shop for its help casting the lead targets used for this detector, Steven Gardiner for his assistance in adapting MARLEY for use with iron and lead, and the authors of Ref.~\cite{wakasa2012} for providing the ${}^{208}\mbox{Pb}(p,n)$ data used for generating predictions with MARLEY.

The COHERENT collaboration acknowledges the Kavli Institute at the University of Chicago for Eljen cell detector contributions. The COHERENT collaboration acknowledges the generous resources provided by the ORNL Spallation Neutron Source, a DOE Office of Science User Facility, and thanks Fermilab for the continuing loan of the CENNS-10 detector. We also acknowledge support from the Alfred~P. Sloan Foundation, the Consortium for Nonproliferation Enabling Capabilities, the National Science Foundation, the Korea National Research Foundation (No. NRF 2022R1A3B1078756), and the U.S. Department of Energy, Office of Science. Laboratory Directed Research and Development funds from ORNL also supported this project. This work was performed under the auspices of the U.S. Department of Energy by Lawrence Livermore National Laboratory under Contract DE-AC52-07NA27344. This research used the Oak Ridge Leadership Computing Facility, which is a DOE Office of Science User Facility. The work was supported by the Ministry of Science and Higher Education of the Russian Federation, Project ``New Phenomena in Particle Physics and the Early Universe'' FSWU-2023-0073.

%\nocite{*}

\bibliography{main}

\clearpage\newpage

\end{document}